\documentclass[aps,twocolumn,amsfonts,amsmath,amssymb,prb]{revtex4}

\usepackage{graphicx}
\usepackage{graphics}
\usepackage{amsmath}
\usepackage{amsfonts}
\usepackage{color}

\begin{document}
%\title{Modelling supercapacitors: Are constant charge surfaces electrodes?}
\title{Structural Transitions at Ionic Liquids Interfaces}

\author{Benjamin Rotenberg$^{1,2}$, Mathieu Salanne$^{1,2,3}$}
\affiliation{$^1$Sorbonne Universit\'es, UPMC Univ Paris 06, CNRS, Laboratoire PHENIX, F-75005, Paris, France}
\affiliation{$^2$R\'eseau sur le Stockage Electrochimique de l'Energie (RS2E), FR CNRS 3459, France}
\affiliation{$^3$Maison de la Simulation, USR 3441, CEA - CNRS - INRIA - Universit\'e Paris-Sud -Universit\'e de Versailles, F-91191 Gif-sur-Yvette, France}

%%%%%%%%%%%%%%%%%%%%%%%%%%%%%%%%%%%%%%%%%%%%%%%%%%%%%%%%%%%%%%%%%%%%%%%%%%%%%%%%%%%%%%%%%%%%%
\begin{abstract}
Recent advances in experimental and computational techniques have allowed for an
accurate description of the adsorption of ionic liquids on metallic electrodes.
It is now well established that they adopt a multi-layered structure, and that
the composition of the layers changes with the potential of the electrode. In some
cases, potential-driven ordering transitions  in the first adsorbed layer have
been observed in experiments
probing the interface on the molecular scale
% (Scanning Tunneling Microscopy or Atomic Force Microscopy) 
or by molecular simulations. This perspective gives an overview of
the current understanding of such transitions and of their potential impact on
the physical and (electro)chemical processes at the interface. In particular,
peaks in the differential capacitance, slow dynamics at the interface and
changes in the reactivity have been reported in electrochemical studies. Interfaces
between ionic liquids and metallic electrodes are also highly relevant for their
friction properties, the voltage-dependence of which opens the way to exciting
applications.
%for which strong voltage-dependency have also been observed.  

\end{abstract}

\maketitle

%%%%%%%%%%%%%%%%%%%%%%%%%%%%%%%%%%%%%%%%%%%%%%%%%%%%%%%%%%%%%%%%%%%%%%%%%%%%%%%%%%%%%%%%%%%%%
%\section{Introduction}

%\begin{figure}[ht!]
%\begin{center}
% \includegraphics[width=5cm]{./TOC.pdf}
%\end{center}
%\caption*{TOC graphic.}
%\end{figure}

Solid-liquid interfaces play key role in many processes, such as catalysis
or electrochemical reactions, to mention only chemistry and energy related
applications. Despite their importance, our understanding of the molecular-scale
structure of such interfaces, where all the essential (electro)chemical processes
occur, has long remained limited compared to the case of the corresponding 
pure solid and liquid phases. Probing directly the interface in experiments is indeed
particularly challenging and computer simulations are also more involved 
due to the symmetry breaking in the direction perpendicular to the interface,
which hinders the efficient use of periodic boundary conditions.
Indeed, although simulating a few tens of 
water molecules may be sufficient to investigate the bulk properties of the
liquid~\cite{lin2009b}, a similar number of molecules results in
finite-size effects different from that, physically relevant, 
induced by the presence of the interface.

%in contact with a surface
%of 1~nm$^2$ will form a layer of width inferior to 2~nm~\cite{tazi2012b}.
%Whether the simulated phase is then the bulk or a confined liquid remains an
%open question.

The past 10 years have witnessed the development of 
many experimental techniques which are sensitive to molecular
arrangements at the interface, such as Scanning Tunneling Microscopy (STM),
Sum-Frequency Generation (SFG), Atomic Force Microscopy (AFM), high-energy X-ray
reflectivity (XR) or Surface Force Apparatus (SFA). They probe the structure of the liquid via different observables (vibrations, electron density, resistance to shear, etc), thus providing complementary views of the interface: For example the SFG signal is dominated by the innermost adsorbed layer~\cite{baldelli2008a,penalber2012a,baldelli2013a}, while AFM or XR studies probe several layers of fluid.~\cite{atkin2007a,hayes2015a,mezger2008a,mezger2015a} In parallel,
the access to high performance computers and the development of new
algorithms~\cite{vandevondele2005a,reed2007a} also allowed to simulate more
accurately solid-liquid interfaces, shedding a new light on interfacial
processes such as adsorption. 
For example, the combination of STM and Density Functional Theory
(DFT) calculations demonstrated that water molecules adsorbed at metal surfaces
exhibit a surprisingly rich variety of structures~\cite{carrasco2012a}. Their
arrangement depends on the interplay between the geometry and energetics of
the water-metal interaction and of the
hydrogen bonding between the water molecules, which varies strongly 
from one metal to another (and even from one crystal plane of a given metal
to another) and with the water coverage of the surface. This structuring impacts
the dynamics at the interface~\cite{limmer2013b} and ultimately the kinetics of
electrochemical processes. Due to the range of length and time scales involved,
from the electron transfer event to the local rearrangements of the interfacial fluid,
a full understanding of the water-splitting mechanisms from computer simulations 
will therefore require bridging the gap between {\it ab
initio}~\cite{cheng2012a,nielsen2015a} and classical~\cite{willard2013a}
approaches.

Here we will focus on a particular class of electrolytes, namely
room-temperature ionic liquids (RTILs). They are increasingly used in
electrochemistry, with applications ranging from energy storage (batteries,
supercapacitors) to electrodeposition~\cite{armand2009a}. Since they are made of
ions, their interfacial properties have long been interpreted following the
Gouy-Chapman-Stern theory. However, many of the underlying asumptions are not
valid due the very high density of ions -- an extreme case considering the
absence of solvent in these liquids~\cite{kornyshev2007a,perkin2013a,lee2015a}. 
A significant number of
experiments and molecular simulations have thus been devoted to the study of the
interfaces of ionic liquids with a solid~\cite{fedorov2014a,hayes2015a}. The main conclusion
arising from XR, AFM, SFA and molecular dynamics (MD) is that the
structure perpendicular to the interface is characterized by a strong layering
of the liquid~\cite{hayes2015a,mezger2008a,perkin2012a,merlet2013c}, as expected
for a molecular liquid, which extends up to a few nanometres. 
The local composition of the layers mostly depends on
the surface charge of the solid~\cite{ivanistsev2014b} and displays strong local correlations due to
charge-ordering.
%\revision{[Il faudra parler soit ici, soit plut\^ot en conclusion, du papier de
%David avec son mod\`ele sur r\'eseau. Je ne sais pas s'il est d\'ej\`a paru,
%mais il le sera sans doute d'ici-l\`a, en attendant on peut toujours citer
%arXiv.]}

Many recent studies on interfaces of RTILs reported intriguing results,
highlighting the role of the molecular structure within the
adsorbed layers. As pointed out in an editorial by Kornyshev and Qiao, it
is indeed necessary to account for the three-dimensionality of the
interface~\cite{kornyshev2014a}. In particular, the formation of an ordered
layer of ions has been reported at the interfaces of
1-butyl-3-methylimidazolium-hexafluorophosphate ([C$_4$mim$^+$][PF$_6^-$]) with
mica~\cite{liu2006a} or with vapor~\cite{jeon2012a}. 
 At electrochemical
interfaces, the contact between RTILs with an electrified metal opens the way
to voltage-induced ordering
transitions within the adsorbed liquid. 
The universality of such
transitions is far from being established, in particular the extent of
concerned RTILs-substrate combinations should be clarified. 
A first objective of this perspective article is therefore
to summarize the studies, both experimental and theoretical, in which such
transitions have been observed. We then discuss the impact of this finding
on the physico-chemical properties of the interface. In particular, the
following questions will be addressed: How can we detect structural transitions in
experiments and in simulations? Is there a templating action from the solid? What
is the main electrochemical signature of these transitions? Is there an impact
on the friction properties of the interface?  Some of these questions remain
open and call for further studies. 

%\section{Structural transitions at aqueous interfaces}
%\subsection{Water at metallic surfaces}
%\subsection{Water in confinement}

%%%%%%%%%%%%%%%%%%%%%%%%%%%%%%%%%%%%%%%%%%%%%%%%%%%%%%%%%%%%%%%%%%%%%%%%%%%%%%%%%%%%%%%%%%%%%
\section{Evidences for structural transitions at ionic liquid interfaces}
\subsection{Experimental studies}

To our knowledge, the first studies dealing with interfacial phase transitions
in Coulomb fluids were conducted by Freyland {\it et al.}~\cite{freyland2008a}.
Their {\it in situ} STM study of the interface between the
[C$_4$mim$^+$][PF$_6^-$] and the (111) face of gold reported the formation of
Moir\'e-like patterns at potentials greater than -0.2~V with respect to a
platinum reference electrode. These were attributed to the formation of an
ordered adlayer of PF$_6^-$. At negative potentials, the STM images were
consistent with the formation a layer of anions with the
($\sqrt{3}\times\sqrt{3}$) structure, indicating a two-dimensional ordering
transition at this interface. It is worth noting that these
observations closely follow a previous work performed on the adsorption of
iodine from aqueous solutions on similar gold surfaces~\cite{tao1992a}. A
further study by the same authors on the electrodeposition of Cd on Au(111) in a
chloroaluminate ionic liquid has also revealed the formation of an ordered
AlCl$_4^-$ adlayer~\cite{pan2007a}. 

\begin{figure}[ht!]
\begin{center}
 \includegraphics[width=0.5\textwidth]{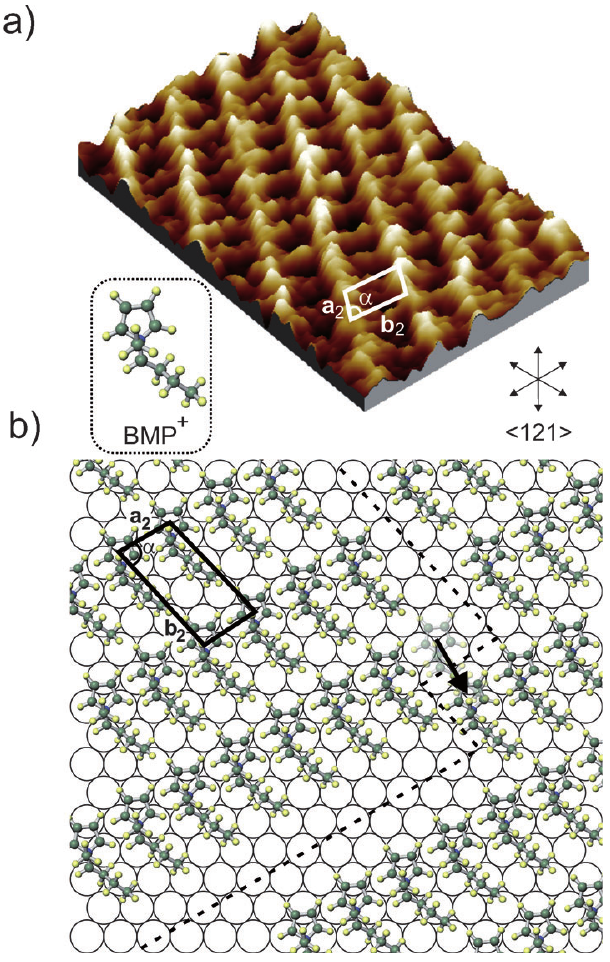}
\end{center}
\caption{a) High-resolution STM image (5.4 nm$\times$7.5 nm) of the BMP$^+$ adlayer on Au(111) at $-$1.4 V and b) proposed structural model, showing two domains of the ($\sqrt{3}\times\sqrt{13}$) BMP$^+$ adlayer separated by a translational domain boundary (indicated by dashed line). Reproduced with permission from reference \citenum{wen2015a}; Copyright: Wiley, 2015.}
\label{fig:stm}
\end{figure}

When changing both the nature of the anion (from PF$_6^-$ to BF$_4^-$) and of
the surface of gold in contact with the RTIL (from (111) to (100)), Su {\it et
al.} have also evidenced the existence of potential-driven ordering
transitions~\cite{su2009a}. Increasing the potential from $-$0.3~V, an ordered
layer of anions is formed between $-$0.1 and 0.4~V. On the contrary, when
scanning in the negative potentials direction, they first observed a loose
film-like layer which was attributed to a disordered adsorption of C$_4$mim$^+$
cations. Then, for potentials lower than $-$0.95~V, perpendicularly oriented
double-row strips were observed. These strips were assigned to the formation of
micelle-like arrangements of aligned C$_4$mim$^+$ cations. These structures also
formed with PF$_6^-$ and SO$_3$CF$_3^-$ anions, but not on (111) surfaces of
gold~\cite{su2009a}, which shows that in this case a structural commensurability
of the adsorbed layer and the metal surface is necessary for the formation of
ordered structures. 

This conclusion was confirmed in a study using {\it in situ} video-STM 
to probe the (111) Au interface with a RTIL composed of a different cation,
namely 1-butyl-1-methylpyrrolidinium (BMP$^+$) associated with the
bis(trifluoromethylsulfonyl)imide anion (TFSI$^-$)~\cite{wen2015a}. Stable
images could only be obtained for negative potentials below $-$1~V. For such
potentials the images showed the formation of ordered structures. Several
distinct arrangements of cations were proposed in order to interpret the
observations at various potentials; one of them is shown in \ref{fig:stm}.
% The main difference between C$_4$mim$^+$ and BMP$^+$ cations is their relative
%dimensions, which confirms the importance of commensurability with the surface. 
 In both structures proposed by Wen {\it et al.}, the cation rings are adsorbed on the surface; in contrast,
the alkyl chains lie flat on the surface only for the lower charge density
(hence lateral cation density in the adsorbed fluid) and extend into the
perpendicular direction for the higher density. The lattice parameters for the
adlayer superstructure decrease accordingly and may change symmetry, resulting
for the densest packing in a square lattice which differs from the
hexagonal substrate.
 Finally, the video-STM furter allowed the first
direct observation of the dynamical evolution of the adsorbed liquid. In
particular, it was found that the fluctuations occur mainly at the boundaries
between ordered domains.

All the structural transitions observed with STM have so far involved Au
electrodes because this metal can be produced as single crystals with
well-defined surfaces. Elbourne {\it et al.} have recently used another
technique, {\it in situ} amplitude-modulated AFM, to study interfaces of highly
ordered pyrolytic graphite (HOPG) instead~\cite{elbourne2015a}. This substrate
presents the advantage of having flat surfaces with high area, and to avoid the
surface reconstructions or etching which can occur in the case of
gold~\cite{su2009a}. These authors studied the effect of applied potential on
the adsorbed layer structure for a [C$_4$mim$^+$][TFSI$^-$] ionic liquid. At the
open-circuit potential, well-defined rows are present on the surface. Unlike
previous works, in which the ordered structures were apparently formed of only
one type of ions, the unit cell is composed of an anion-cation-cation-anion
arrangement~\cite{elbourne2015a}. This structure changes markedly with surface
potential or when relatively low concentrations of lithium or chloride ions are
present in the RTIL.
  %The changes induced by the addition of small amounts of [Li$^+$[[TFSI$^-$] or [C$_4$mim$^+$][Cl$^-$] were also probed because Li$^+$ and Cl$^-$ are often found as impurities in RTILs. The   

The variety of systems in which transitions are observed clearly show that it is
a common feature of metal/ionic liquids interfaces. However, the few works
reported so far raise very interesting questions. In particular, all the ordered
structures proposed so far to interpret the STM data are composed
of a single species only. However, the relatively small applied potentials which
are used ($\sim\pm$1~V) may not be sufficent to fully separate cations from
anions. In the case of the AFM study, the anion-cation-cation-anion rows result
in an overall neutral layer but there is a strong charge imbalance on the
nanometre scale. A few hypotheses can thus be proposed to explain the
observations: i) there may be a specific absorption of the ions on gold with the
formation of partially covalent bonds~\cite{anderson2014a} ii) on top of the
observed layer, there could be an oppositely charged layer of ions which is not
observed by the experiment and iii) it remains possible that the proposed
structures, which are only based on the relative size of the ions (keeping the
possibility of some kind of conformational ordering, for example only the
imidazolium rings of the cations would lie parallel to the surface), are not the
correct ones. Using additional techniques such as SFG, which is sensitive to the
orientation of the ions~\cite{baldelli2013a}, could possibly shed a
complementary light on this issue.   
Another open question is whether commensurability between the adsorbed layer and
the metal substrate is necessary to observe a transition. Here also, the recent
study performed with HOPG electrodes~\cite{elbourne2015a} suggests that it is
not the case, and that the electric fluctuations at an homogeneous and flat
metallic surface are sufficient to trigger ordering transitions in the interfacial layer of RTIL. In addition, the use of carbon electrodes in this study demonstrates that a perfect metallic behavior is not necessary to induce such transitions.

\subsection{Computer simulations}

\begin{figure*}[ht!]
\begin{center}
 \includegraphics[width=\textwidth]{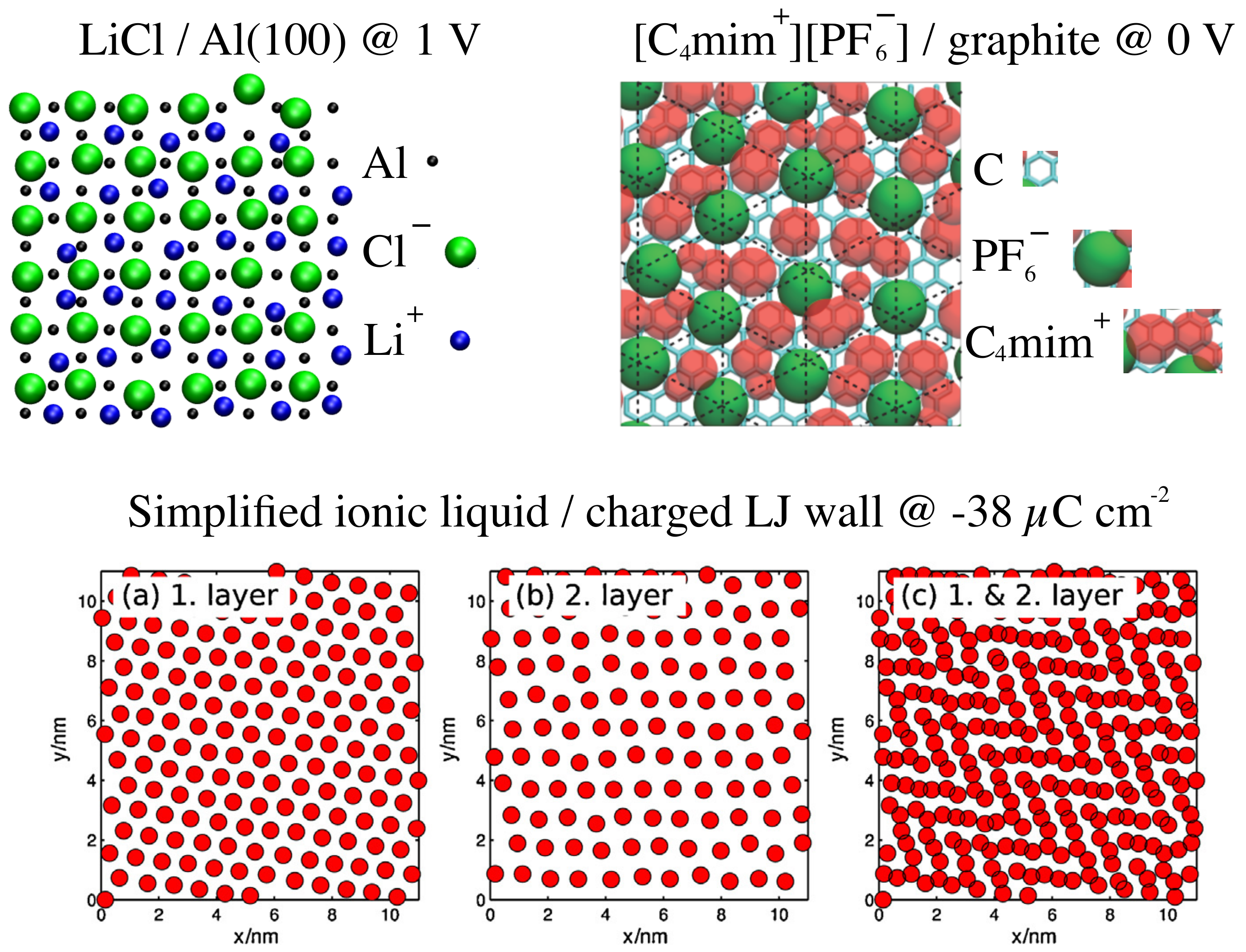}
\end{center}
\caption{Snapshots of typical ordered structures observed in computer
simulations. Top left: First adsorbed layer of a LiCl molten salt electrolyte on
the (100) surface of an aluminum electrode at a negative potential. Adapted from
reference~\citenum{tazi2010a}. Top right: First adsorbed layer of a
[C$_4$mim$^+$][PF$_6^-$] RTIL on a graphite electrode at a neutral potential.
Adapted from reference~\citenum{merlet2014a}. Bottom: First two adsorbed layer
of a simplified RTIL on a charged Lennard-Jones wall with a large negative surface charge density.   
Reprinted from reference~\citenum{kirchner2013a}, Copyright 2013, with permission from Elsevier.
}
\label{fig:snapshots}
\end{figure*}

In order to simulate electrochemical systems, it is necessary to fix the
potential of the electrode. In classical molecular dynamics or Monte-Carlo, this
can be done by various methods~\cite{merlet2013c}. Our approach consists in
treating the partial charges carried by the electrode atoms as additional
degrees of freedom which fluctuate during the simulation. Their values are
determined at each time step from a self-consistent
calculation~\cite{siepmann1995a,reed2007a,merlet2013b}. In such simulations, the
electrochemical cell consists in a wide slab of electrolyte held between two
electrodes with different voltages. Like in experiments, the potentials are not
absolute. The only fixed quantity is the potential difference between the two
electrodes $\Delta\Psi=\Psi^+-\Psi^-$, although it is also possible to calculate
the potential of each electrode with respect to the bulk liquid in the case of
flat electrodes; we will note this potential $\Psi^{\rm elec/bulk}$. In the
following, we will assimilate an ordering transition to an abrupt change in the structure observed when changing the potential. However, it is worth noting that first-order transitions are associated with a discontinuity in an order parameter and a corresponding singularity in a partition function, which are not easy to prove in simulations~\cite{chandler-livre}. This point will be further discussed in the next section.

Using this simulation approach, a first example of voltage-driven transition was
reported for a rather exotic system, formed with a high temperature molten salt
(LiCl) and an aluminum electrode with its (100) surface in contact with the
liquid~\cite{pounds2009b,tazi2010a}. An advantage of this system is that a
polarizable force field could be built directly from accurate DFT calculations
using a generalized force-matching approach~\cite{pounds2009b}.For potential
drops $\Psi^{\rm elec/bulk}$ across the interface more negative than $-$1.76~V,
which corresponds to the point of zero charge (PZC), the molten salt adopted a
disordered structure at the interface, while for larger potentials an ordered
structure was obtained~\cite{pounds2009b}. This structure, which is shown on the
top panel of \ref{fig:snapshots}, was commensurate with the aluminum substrate,
and a strong alignment of the dipole components of the chloride anion and the
normal of the surface was observed for large potentials~\cite{tazi2010a}.
Interestingly, no transition was observed when the plane of the metal was
changed to (110) instead of (100), but a different ordered structure was then
obtained showing that an epitaxial mechanism is at play, whereby the molten
salt adapts it structure to that of the electrode surface.

In a recent work, Kirchner {\it et al.} studied interfaces between primitive
models of ionic liquids and solid surfaces with various net charges ({\it i.e.}
the electric potential was not controlled)~\cite{kirchner2013a}. At certain
charge densities ($\sim$~$-16$~$\mu$C cm$^{-2}$) the structure of the adsorbed
layer of cations undergoes a structural transition to a surface-frozen monolayer
of densely packed counter-ions with a Moir\'e-like structure. At lower surface
charge densities ({\it i.e.} lower than $-30$~$\mu$C cm$^{-2}$), they even
observed the formation of an herring-bone structure arising from the
superposition of two ordered monolayers of ions (see the bottom panel of 
\ref{fig:snapshots}). These findings provide an interesting support for
the STM studies discussed above, but it is worth noting that the charge
densities employed are somewhat larger than the experimental ones -- they would
correspond to potentials  which are above the electrochemical window of typical
RTILs.

Going towards more realistic models, an ordering transition was reported from
molecular simulations for the interface between [C$_4$mim$^+$][PF$_6^-$] (for
which a coarse-grained force field was used) and an electrified surface of
graphite~\cite{merlet2014a}. The presence of the ordered structure could be
monitored by computing the in-plane structure factor in the first layer of the
adsorbed liquid. This structure factor was liquid-like on a wide range of
potentials, but it showed some strong Bragg-like peaks suggesting a
2-dimensional lattice-like organization for both the anions and cations, which
is shown in the top-right panel of \ref{fig:snapshots}. This ordered structure
contained on average as many anions as cations, and it was also observed by
Kislenko {\it et al.} in simulations of the same RTIL (with an all-atom model)
adsorbed on an uncharged surface of graphite~\cite{kislenko2009a}. By using
importance sampling techniques, it could be shown by Merlet {\it et al.} that
this structure was the most stable one for small positive potentials
($\sim0<\Psi^{\rm elec/bulk}<\sim0.5$~V) and metastable for small negative
electrode potentials. It is worth noting that similarly to the experimental work
of Elbourne {\it et al.} involving HOPG electrodes, no commensurability with the
electrode surface seems necessary to observe such ordered structures. 

So far, no ordering transitions have been observed using more elaborate, all-atom models of RTILs in contact with electrodes at fixed potential. In particular, the adsorption of [C$_4$mim$^+$][PF$_6^-$] and [C$_4$mim$^+$][BF$_4^-$] on electrified surfaces of gold was studied by Hu {\it et al.},~\cite{hu2013a} but they did not report any ordering transition similar to the experimental observations by STM. 

These simulation results, while confirming the possibility of transitions in the
adsorbed layer of the fluid, also open their share of questions. Future works
will need to address the issue of finite-size effects, since there must be a
commensurability between the formed ordered structure and the simulation cell.
Timescales are important too, since metastable states may be much longer-lived
than the typical simulation times, which are on the order of the nanosecond only
due to the computational cost. There is therefore a possibility that the
reported transitions are artefacts of the simulation setups, but the
similarities with experimental findings seem to weaken this hypothesis. The
question of specific interactions with surfaces such as gold will also have to be
treated. This requires in turn the development of accurate force fields for this purpose. 
First steps in this direction have recently been made in the case of carbon
materials~\cite{pensado2014a}.

%%%%%%%%%%%%%%%%%%%%%%%%%%%%%%%%%%%%%%%%%%%%%%%%%%%%%%%%%%%%%%%%%%%%%%%%%%%%%%%%%%%%%%%%%%%%%
\section{Impact of the transitions on physico-chemical properties}

We now turn to the consequences of structural transitions within the adsorbed
fluid on the physico-chemical properties of the interface. Specifically, we
discuss the impact of voltage-induced transitions on the electrochemical
response of the electrode-RTIL interface in terms of differential capacitance,
cyclic voltammograms and electrochemical reaction, as well as on the mechanical
response (solid-liquid friction). 

\subsection{Peaks in the differential capacitance}

The differential capacitance $C_{\rm diff}$ measures the response of the average 
surface charge density $\left\langle\sigma\right\rangle$ 
to changes in the voltage $\Delta\Psi$:
\begin{align}
C_{\rm diff} = \frac{ \partial \left\langle\sigma\right\rangle }
{ \partial \Delta\Psi }
\; .
\label{eq:capa1}
\end{align}
By definition, a capacitor corresponds to a voltage-independent differential
capacitance. However, the charge of the electrode reflects the composition and
the charge distribution within the interfacial liquid. As a result, one should
expect a signature of abrupt structural changes at voltages corresponding to
putative phase transitions in the electrode charge, hence peaks in the
corresponding differential capacitance. While experimentally such peaks have
indeed been observed~\cite{su2009a,cannes2013a,costa2015a}, their possible link with changes in the
structure or the interface has been difficult to demonstrate until recently, due
to the experimental challenges of {\it in situ} imaging and the occurence of other
processes such as surface reconstruction of the electrode.

%\revision{[Parler de Baldelli?]}

Indeed, in their STM study of [C$_4$mim$^+$][BF$_4^-$] on a (100) gold electrode, Su {\it et al.} have also measured the capacitance of the interface. They observed a fivefold increase in this quantity in the potential region of transition
from anion adsorption to cation adsorption.
%This study further demonstrated the surface-etching of the gold electrode
%in the cation adsorption regime, which complicates the microscopic 
%interpretation of the peak in differential capacitance. 
In the case of the {\it in situ} video-STM study of the 
[BMP$^+$][TFSA$^-$] on a (111) gold electrode~\cite{wen2015a}, 
thanks to the high temporal resolution or the video-STM technique, 
the authors were able to visualize the evolution of the interfacial fluid 
during cyclic voltammetry (CV) experiments. % at a scan rate of 10~mV.s$^{-1}$.
The cyclic voltammogram displays two current peaks associated 
with two surface transitions which could also be linked to the formation of the
ordered cationic structures, such as the one shown on \ref{fig:stm}, upon increasingly negative surface
charge density. 
%In both structures, the cation rings are adsorbed on the surface; in contrast,
%the alkyl chains lie flat on the surface only for the lower charge density
%(hence lateral cation density in the adsorbed fluid) and extend into the
%perpendicular direction for the higher density. The lattice parameters for the
%adlayer superstructure decrease accordingly and may change symmetry, resulting
%for the densest packing in a square lattice which differs from the
%hexagonal substrate.

% Finally, the video-STM furter allowed the first
%direct observation of the dynamical evolution of the adsorbed liquid. In
%particular, it was found that the fluctuations occur mainly at the boundaries
%between ordered domains.

\begin{figure*}[ht!]
\begin{center}
 \includegraphics[width=0.75\textwidth]{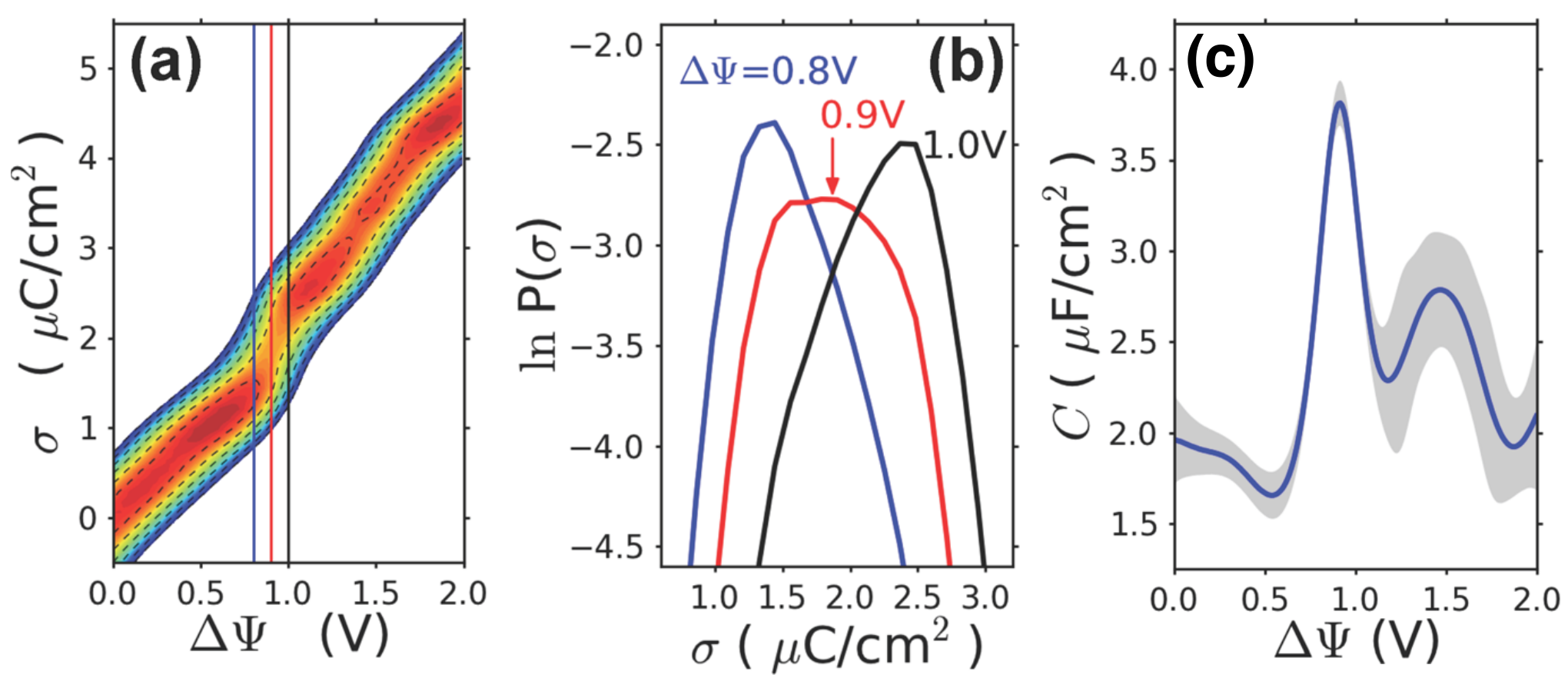}
\end{center}
\caption{(a) Calculated probability distribution of the charge density $\sigma$
of graphite electrodes in contact with the [C$_4$mim$^+$][PF$_6^-$] ionic liquid
with respect to the applied potential $\Delta \Psi$. The two-dimensional graph
of the distribution employs a logarithmic scale with lines separated by a
difference of 0.5 and is plotted as a function of $\sigma$ in (b). Note the fat tails in the distribution, $P(\sigma)$ and the markedly nonlinear shifts with changing voltage. (c) Differential capacitance, $C$, as a function of $\Delta \Psi$. Adapted from reference~\citenum{merlet2014a}.}
\label{fig:capa}
\end{figure*}

In computer simulations, it is relatively straightforward to calculate the
capacitance of the interface in constant potential simulations.  The generic
method consists in simulating an electrochemical cell at various voltage and
extracting the average surface charge. Then the
$\left\langle\sigma\right\rangle=f(\Delta\Psi)$ plot is differentiated, which
provides the differential capacitance through \ref{eq:capa1}. However, close
to a transition, a large peak in the capacitance is expected, so that many
voltages should in principle be sampled in this region. 
An alternative was
recently proposed, which consists in using importance sampling
methods~\cite{limmer2013a,merlet2014a}. In short, by using the whole
distribution of surface charges during the simulations, it is possible to sample
the probability distributions of any variable as continuous functions of the
applied potential. There is in principle no need to acquire more data close to
the transition, the only requisite is to have a good overlap between the
histograms of surface charges from the various voltages.

The probability distribution of the charge density $\sigma$ of graphite
electrodes in contact with the [C$_4$mim$^+$][PF$_6^-$] ionic
liquid~\cite{merlet2014a} obtained with this approach is shown on
\ref{fig:capa}(a). The figure shows the probability distribution on a logarithmic
scale. It is clear that there are three branches along which the distribution of
the surface charge distribution shifts almost linearly upon increasing the potential.
% at 0.6, 1.2 and 1.8~V, which correspond to particularly
%favorable configurations for the interfacial fluid, 
%separated by saddle points. 
These branches are separated by more complex changes in the distribution around
particular voltages.
We will focus on the one occuring at $\Delta \Psi$~=~0.9~V since
this is the potential for which the order-disorder transition discussed above
occurs. \ref{fig:capa}(b) shows the distribution $P(\sigma)$ at three
applied voltages (0.8~V, 0.9~V and 1.0~V). They are characteristic of a
first-order phase transition. Away from phase coexistence (at 0.8~V and 1.0~V),
they show the presence of ``fat tails'', which are due to the small probability
of obtaining the metastable phase. At the transition the distribution displays
hints of bimodality, which is expected if the two phases are equiprobable. 
However it would be necessary to simulate larger systems to fully conclude on this point.

 The differential capacitance computed from these simulations is shown on
\ref{fig:capa}(c). Note that much better statistics could again be
obtained compared to the usual method involving \ref{eq:capa1} by using the Johnson-Nyquist relation,
\begin{equation}
C=\frac{S}{k_{\rm B}T}\left\langle(\delta\sigma)^2\right\rangle,
\end{equation}
\noindent where $S$ is the surface of the electrode and $\delta\sigma=\sigma-\langle\sigma\rangle$ are the fluctuations in the electrode surface charge density. A large peak in the differential capacitance is observed at the applied voltage where the transition occurs, which is consistent with the experimental findings of Su {\it et al.} Note that again, larger systems should lead to a singular charge-density transition in a macroscopic limit~\cite{chandler-livre}. Our simulations therefore confirm that the presence of large peaks in experimental measures of the capacitance of an interface can indeed be the signature that a potential-driven transition is occuring.

\subsection{Hysteresis and slow dynamics}

The above-mentioned domain boundaries between phases also have important 
implications by themselves, due to the entailed free energy cost. 
In three dimensions, this would be a surface free
energy. In the present case the topology of the boundary between interfacial
domains at the surface of the electrode remains to be clarified~\cite{kornyshev2014a}.
As a result, annealing these boundaries, either between grains of otherwise 
identical domains, or between different domains, requires overcoming the
corresponding free energy barrier. In practice, the consequences of these barriers 
are observable as long time scales in the dynamics of the interface or as
hysteresis in cyclic voltammetry. 

Uysal \emph{et al.} reported a potential-dependent hysteresis at an electrified
graphene/RTIL interface~\cite{uysal2014a}. X-ray reflectivity
measurements during cyclic voltammetry and potential step measurements are
used to probe the electronic density in the direction perpendicular to an epitaxial
graphene surface, within the adsorbed  [C$_9$mim$^+$][TFSI$^-$] ionic liquid.
The resulting profiles were consistent with that obtained from MD simulations,
by assuming a combination of two limiting structures, with weights varying as a
function of applied voltage. The structure evolves very slowly after a potential
step, with processes occuring over time scales exceeding 10~s. In addition, the
CV scans exhibit significant (scan rate dependent) hysteresis. While in this
work the authors safely indicated that the nature of the apparent barrier and
the associated mechanism require further investigation, these observations
clearly point to the crucial role of structural transitions and the associated
domain boundaries in the observed hysteresis and slow dynamics. Another manifestation of slow processes occuring at the ionic liquid / electrode interface was reported by Roling {\it et al.}, who have carefully analysed the capacitance spectra on a broad range of frequencies~\cite{roling2012a,druschler2012a}. Although no particular ordering transition was observed by STM, these authors concluded that the slower capacitive process could be related to structural reorganisations of the gold surface or to strong rearrangements in the first adsorbed layer of ions.

Recently, Limmer proposed a detailed study of these effects using 
a coarse-grained model capturing strong inter-ionic
correlations~\cite{limmer2015a}. Its limited computational cost compared to
molecular simulations allowed for a systematic finite-size scaling analysis,
which demonstrated the first-order nature of the 
fluctuation-induced transition and spontaneous charge density
ordering at the interface, in the presence of an otherwise disordered bulk
solution, already observed in molecular simulations~\cite{merlet2014a}. 
A crucial step in this demonstration is the extensive growth of the free energy
barrier between phases analogous to the ones observed in reference~\citenum{uysal2014a},
which indeed implies hysteresis and long time scales.

\subsection{Impact on reactivity}

Structural changes in the ionic liquid at the interface also have implications 
on the local environment of other species in the liquid, in particular
electro-active species. This in turn may result in changes in their reactivity.
A direct observation of this feature has recently been reported by 
Garcia-Rey and Dlott, who studied CO$_2$ reduction on a
polycrystalline Ag electrode, with 1-ethyl-3-methylimidazolium tetrafluoroborate
[C$_2$mim$^+$][BF$_4^-$] containing 0.3~mol\% water as electrolyte~\cite{garciarey2015a}. 
Such systems have been shown to reduced the overpotential for CO$_2$ reduction.
SFG and IR were used to probe the surface field experienced by the 
adsorbed CO molecules produced by the electrochemical reduction of CO$_2$.
From the CO Stark shift, a sudden increase of the field at the electrode surface
was observed at the threshold potential for CO$_2$ reduction, which could be
traced back to a structural transition within the RTIL -- even though no
information could be obtained on the nature of these structural changes.
Nevertheless, this study illustrates the potential benefit of exploiting 
the peculiar structure of ionic liquid interfaces 
and the voltage-driven changes thereof (with potentially much greater diversity
than in solvent-based electrolytes) for electrochemical reactions.

%\revision{Parler de Marcus et des PMFs: Bedrov, Varella, etc}

\subsection{Voltage-dependent friction}

\begin{figure}[ht!]
\begin{center}
 \includegraphics[width=0.5\textwidth]{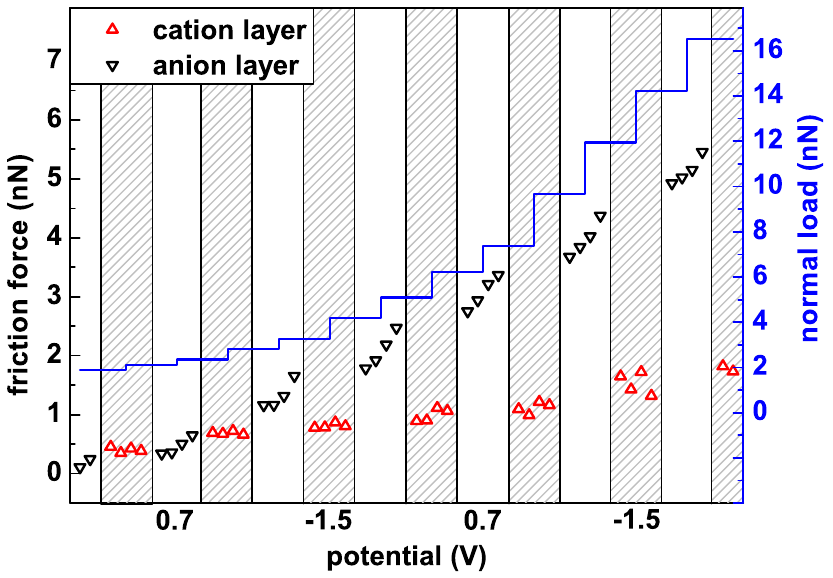}
\end{center}
\caption{Friction forces as functions of load and applied potential recorded for a sharp AFM tip sliding on Au(111) in [BMP$^+$[FAP$^-$]. Each data point represents the average over a full scan frame of 14 nm side length. The potential is switched between $-$1.5 and 0.7~V for each value of the applied normal load. Reprinted with permission from reference \citenum{sweeney2012a}; Copyright 2012 by the American Physical Society.}
\label{fig:friction}
\end{figure}

Finally, voltage-driven changes in the structure and composition of the
interfacial fluid also have implications from the dynamical point of view.
Sweeney {\emph et al.} conducted nanotribology experiments to probe the
lubrication properties of 1-butyl-1-methylpyrrolidinium tris(pentafluoroethyl)
trifluorophosphate ([BMP$^+$][FAP$^-$]) confined between silica colloid probes 
or sharp silica tips and a Au(111) substrate, using AFM~\cite{sweeney2012a}.
As the composition of the adsorbed layers is tuned by the electrode potential,
from cation-enriched to anion-enriched, the friction also evolves.
\ref{fig:friction} illustrates that
these variations are directly linked to the nature of the sliding plane, 
which may correspond to cation or anion layers, depending on the electrode 
potential and on the normal load exerted on the confined fluid. 
 While the voltage-driven structural changes on the
microscopic interfacial structures remain to be investigated, such studies
open the way to a new tuning of frictional forces at the molecular scale without
changing the substrate.

More detailed information on the role of key microscopic and macroscopic
factors can be obtained using molecular simulations,
such as load, shear velocity, surface topology and length of alkyl side chains
in the ionic liquid~\cite{mendonca2013a}. Simulations with fixed surface charge
density (instead of potential) have further evidenced two mechanisms underlying
friction changes in such systems, namely charge effects on normal and in-plane
ordering in the film, as well as swapping between anion and cation layers at the
surface~\cite{fajardo2015a}. 

%\revision{Je ne connaissais pas le papier de Fernando, Alexei et Urbakh, ils
%auraient quand m\^eme pu citer le JPCC de l'an dernier...
%Electrotunable Lubricity with Ionic Liquid Nanoscale Films,
%O. Y. Fajardo,	F. Bresme, A. A. Kornyshev \& M. Urbakh, 
%Scientific Reports, 5, 7698, 2015 (doi\string:10.1038/srep07698)
%}

\section{Summary and outlook}

There is now a large body of experiments pointing towards the existence of
potential-driven transitions at the interface between ionic liquids and metallic
electrodes. However, as discussed above, the exact structure and composition of
the ordered phases remain open questions. Computer simulations bring some
theoretical support on the question, but they are still scarce because of the
technical difficulty associated with the use of constant applied potential ensemble. They also suffer from sampling issues (both in size and time) which render the observation and the characterization of the transitions difficult. The current works, in which the interactions are determined using classical force fields may also be limited if particular bonding occurs at the interface. The recent inclusion of constant voltage methods in DFT-based molecular dynamics packages~\cite{golze2013a} may open new opportunities for tackling this difficult problem, as it was shown recently in the context of nanotribology.~\cite{page2014a,li2015e}

In addition, the role of many parameters remain to be established. For example,
how do the composition of RTIL and the possible presence of impurities affect the occurence of ordering transitions? What is the impact of the temperature of the systems? Also many applications of RTILs use them in the presence of a solvent, which will also impact the whole structure of the electric double layer. Finally, although it is clear that the nature of the substrate plays a strong role, it is not sure that there is always a commensurability between the ordered structure of the liquid and the metal. Topological defects at the surface of the metal may also play a predominant role,~\cite{black2015a} and it is likely that corrugation effects can modify the formation and/or the detection of ordered layers: Recent simulations have shown that the heterogeneous nucleation of ice at a surface depended markedly on the morphology of the latter.~\cite{fitzner2015a} Additional works with varying metal electrodes will allow to better understand these issues. 

Whether these transitions will have practical applications remains an open
question, but they clearly impact a lot the physico-chemical properties: peaks
in the differential capacitances, slow dynamics at the interface, varying
reactivity and voltage-dependent friction properties have already been reported. 
Overall such transitions reinforce the view of RTILs as solvents with
multifaceted properties, with a composition that can be specifically tailored to a given task.

\providecommand*\mcitethebibliography{\thebibliography}
\csname @ifundefined\endcsname{endmcitethebibliography}
  {\let\endmcitethebibliography\endthebibliography}{}

%\bibliography{references}

\end{document}